# Direct Multipoint Observations Capturing the Reformation of a Supercritical Fast Magnetosonic Shock

D. L. Turner[1], L. B. Wilson III[2], K. A. Goodrich[3], H. Madanian[4], S. J. Schwartz[5], T. Z. Liu[6,7], A. Johlander[8], D. Caprioli[9], I. J. Cohen[1], D. Gershman[2], H. Hietala[10], J. H. Westlake[1], B. Lavraud[11,12], O. Le Contel[13], and J. L. Burch[4]

[1]The Johns Hopkins University Applied Physics Laboratory, Laurel, MD, USA
[2]NASA Goddard Space Flight Center, Greenbelt, MD, USA
[3]Department of Physics and Astronomy, West Virginia University, Morgantown, WV, USA
[4]Southwest Research Institute, San Antonio, TX, USA
[5]Laboratory for Atmospheric and Space Physics, University of Colorado, Boulder, CO, USA
[6]Cooperative Programs for the Advancement of Earth System Science, University Corporation for Atmospheric Research, Boulder, CO, USA.
[7]Geophysical Institute, University of Alaska, Fairbanks, Fairbanks, AK, USA.
[8]Department of Physics, University of Helsinki, Helsinki, Finland
[9]Department of Astronomy and Astrophysics, University of Chicago, Chicago, IL, USA
[10]The Blackett Laboratory, Imperial College London, UK
[11]Laboratoire d'Astrophysique de Bordeaux, Univ. Bordeaux, CNRS, B18N, Pessac, France
[12]Institut de Recherche en Astrophysique et Planétologie, CNRS, UPS, CNES, Université de Toulouse, Toulouse, France
[13]Laboratoire de Physique des Plasmas, UMR 7648, CNRS/Ecole Polytechnique IP Paris / Sorbonne Université / Université Paris Saclay / Observatoire de Paris, Paris, France

## Abstract

Using multipoint Magnetospheric Multiscale (MMS) observations in an unusual string-of-pearls configuration, we examine in detail observations of the reformation of a fast magnetosonic shock observed on the upstream edge of a foreshock transient structure upstream of Earth's bow shock. The four MMS spacecraft were separated by several hundred km, comparable to suprathermal ion gyro-radius scales or several ion inertial lengths. At least half of the shock reformation cycle was observed, with a new shock ramp rising up out of the "foot" region of the original shock ramp. Using the multipoint observations, we convert the observed time-series data into distance along the shock normal in the shock's rest frame. That conversion allows for a unique study of the relative spatial scales of the shock's various features, including the shock's growth rate, and how they evolve during the reformation cycle. Analysis indicates that: the growth rate increases during reformation, electron-scale physics play an important role in the shock reformation, and energy conversion processes also undergo the same cyclical periodicity as reformation. Strong, thin electron-kinetic-scale current sheets and large-amplitude electrostatic and electromagnetic waves are reported. Results highlight the critical cross-scale coupling between electron-kinetic- and ion-kinetic-scale processes and details of the nature of nonstationarity, shock-front reformation at collisionless, fast magnetosonic shocks.

# 1. Introduction

Collisionless, fast-magnetosonic shocks are ubiquitous features of space plasma throughout the Universe [e.g., Kozarev et al., 2011; Ghavamian et al., 2013; Masters et al., 2013; Cohen et al., 2018]. At magnetohydrodynamic (MHD) scales, incident super-fast-magnetosonic plasma slows and deflects across a shock transition region in a manner generally consistent with the Rankine-Hugoniot jump conditions [e.g., Viñas and Scudder, 1986]. Above a critical Mach number, a significant fraction of incident ions must be reflected by the shock front and return back upstream, contributing to the partitioning of energy by the shock and enabling upstream information of the shock itself to propagate throughout the quasi-parallel (i.e., the angle between the incident magnetic field and shock normal direction is less than ~45 deg) foreshock region [e.g., Eastwood et al., 2005; Caprioli et al., 2015]. Finer-scale (i.e., ion and electron kinetic scales) physics are clearly also significant considering the formation of ion-scale structures, such as the magnetic "foot" and "overshoot" on either side of the ramp of supercritical shocks [e.g., Gosling and Robson, 1985], and ion- and electron-kinetic-scale wave modes present around the shock ramp and in both the upstream and downstream regimes [e.g., Wilson et al., 2007; Wilson et al., 2012; Breuillard et al., 2018a-JGR; Chen et al., 2018-PRL; Goodrich et al., 2019].

By their nature, collisionless shocks convert the energy necessary to slow and divert super-fast-magnetosonic flows across a transition region that is much shorter than the collisional mean-free path of particles in the plasma. There is still much debate over the principal physical mechanisms responsible for the bulk deceleration and heating of plasma across the shock [e.g., Wilson et al., 2014a]. Recent results from simulations and observations at Earth's bow shock have highlighted the importance of energy dissipation and heating via ion-kinetic coupling between the incident plasma and reflected ion populations [Caprioli and Spitkovsky, 2014a; 2014b; Goodrich et al., 2019] and via electron-kinetic-scale physics such as energy dissipation in large-amplitude, electron-scale electrostatic waves [Wilson et al., 2014b; Goodrich et al., 2018], whistler-mode turbulence [Hull et al., 2020-JGR], and reconnection along thin, intense, electron-scale current sheets [Gingell et al., 2019; Liu et al., 2020]. Upstream of quasi-parallel supercritical shocks, large-scale transient structures can form in the ion foreshock due to reflected ions' kinetic interactions with the turbulent and discontinuous incident plasma [e.g., Omidi et al., 2010; Turner et al., 2018; Schwartz et al., 2018; Haggerty and Caprioli, 2020]. Often, new fast magnetosonic shocks form on the upstream sides of foreshock transient structures as they expand explosively into the surrounding solar wind and foreshock plasmas [e.g., Thomsen et al., 1988; Liu et al., 2016].

State-of-the-art simulations remain computationally limited and not yet capable of capturing both true electron-to-ion mass ratios and electron plasma to cyclotron frequency ratios in three-dimensions (and thus coupling between those populations is not necessarily accurate). Meanwhile observations are most often limited by single-point observations, resulting in spatiotemporal ambiguity, and/or inadequate temporal resolution. Furthermore, theory and observations [e.g., Morse et al., 1971; Krasnoselskikh et al., 2002; Sundberg et al., 2017; Dimmock et al., 2019; Madanian et al., 2021] indicate that supercritical shocks undergo periodic reformation, also known as nonstationarity, which further complicates discerning details in single-point observations of well-formed shocks. In this study, we examined fortuitous multipoint observations during a single cycle of shock reformation on the upstream edge of a foreshock transient using NASA's Magnetospheric Multiscale (MMS) mission upstream of Earth's bow shock.

# 2. Data and Observations

Data from NASA's MMS mission [Burch et al., 2016a] are utilized for this study. MMS consists of four spacecraft that are identically instrumented to study electron-kinetic scale physics of magnetic reconnection [e.g., Burch et al., 2016b; Torbert et al., 2018]. Here, we use data from the fluxgate [Russell et al., 2016] and search-coil [Le Contel et al., 2016] magnetometers, ion and electron plasma distributions and moments [Pollock et al., 2016], and electric fields [Ergun et al., 2016; Lindqvist et al., 2016]. Typically, the four MMS spacecraft are held in a tight tetrahedron configuration, with inter-satellite separations of ~10 to 100 km. However, during a ~1-month period in 2019, the spacecraft were realigned into a "string-of-pearls" configuration, in which they were separated by up to several 100 km along a common orbit to study turbulence in the solar wind at ion kinetic scales. While in both the tetrahedron and string-of-pearls configurations, MMS are ideal for disambiguating spatiotemporal features in dynamic space plasmas. With this uncommon MMS configuration, we examined in detail a foreshock transient event reported in Turner et al. [2020], which showcased an intriguing evolution of a fast magnetosonic shock.

Figure 1 shows data from the event. Panels a) – g) show data from MMS-1, highlighting the foreshock transient. The transient, associated with the deflection of ion velocity between 04:38:45 and 04:39:28 UT in Fig. 1d, was originally classified by Turner et al. [2020] as a foreshock bubble [e.g., Omidi et al., 2010; Turner et al., 2013], but upon a more detailed investigation for this study, the event may be a hot flow anomaly [e.g., Schwartz et al., 2000]. Evidence supporting this diagnosis consists of the orientation of the associated solar wind discontinuity (normal direction, n = [0.69, -0.51, -0.52]$_{GSE}$), which would have already intersected Earth's bow shock (located ~ 0.5 Earth radii, $R_E$, from MMS at the time), and the orientation of the foreshock transient. More detail on this ion foreshock transient is provided in the next section and appendix material. For the interest of this study, it is irrelevant whether this transient structure was a foreshock bubble or hot flow anomaly, since here we are only concerned with the compression region and formation of a fast magnetosonic shock on the transient's upstream edge.

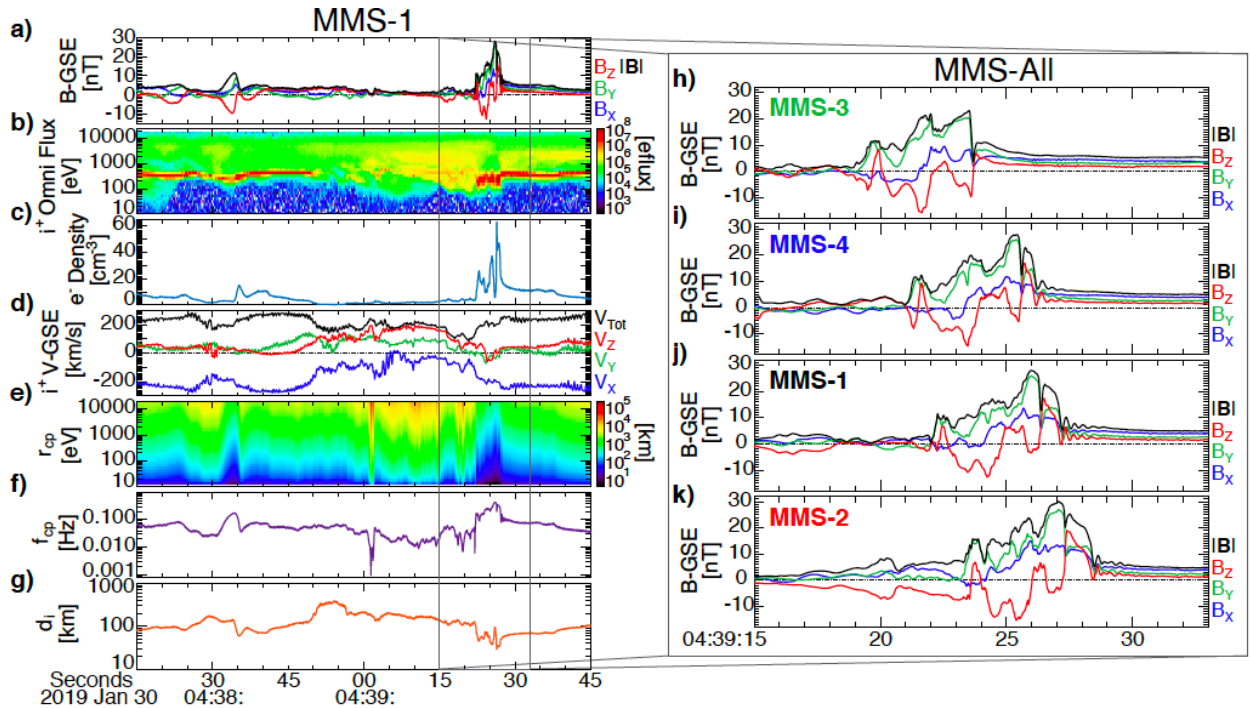

**Figure 1:** Overview of the event observed by MMS. a) – g) show data from the foreshock transient observed by MMS-1 on 30 Jan 2019, including: a) magnetic field vector in GSE coordinates (XYZ in blue, green, and red, respectively) and magnitude (black); b) ion omnidirectional energy-flux (color, units eV/cm²-s-sr-eV); c) electron density; d) ion velocity vector in GSE coordinates (XYZ in blue, green, and red, respectively) and magnitude (black); e) proton gyro-radius (color) as a function of energy and time; f) proton gyro-frequency; g) ion inertial length. h) – k) show magnetic field vectors and magnitudes from all four MMS spacecraft zoomed in on the feature of interest in this study.

Figure 1h) – k) shows magnetic fields observed by all four MMS spacecraft between 04:39:15 and 04:39:33 UT. MMS-3 was the first to pass through the compression region (characterized by the enhanced magnetic field strength and plasma densities) on the upstream side of the foreshock transient, followed next by MMS-4, -1, and finally -2. The four spacecraft observed notable similarities and differences in the structure. All four spacecraft observed large-amplitude waves throughout the compression region; for example, the distinct peaks in |**B**| and corresponding oscillations in the B-field components observed by MMS-3 between 04:39:19 - :24 UT are also evident at the other three spacecraft. However, the differences between the four spacecraft observations at the sharp ramp in magnetic field strength (and density) separating the compression region from the upstream solar wind (e.g., around 04:39:24 at MMS-3) are of interest considering nonstationarity of fast magnetosonic shocks [e.g., Dimmock et al., 2019]. A new compression signature, first observed by MMS-3 at 04:39:24 UT then at MMS-4, -1, and -2 at 04:39:26, :27, and :28 UT, respectively, increases in amplitude and duration on the upstream edge. That was the feature that we focused on in detail for this study.

## 3. Analysis and Results

To properly analyze a shock structure, its orientation and speed must first be established. Using coplanarity analysis [Schwartz et al., 1998] with observations of the ramps in |**B**| observed by all four MMS spacecraft (see appendix), a boundary normal was estimated as [0.54, -0.38, -0.74] ± [0.10, 0.10, 0.10] in GSE coordinates. Comparing that normal direction to the upstream B-field, [1.94, 1.16, 0.30]$_{GSE}$ nT, the foreshock transient's shock was in a quasi-perpendicular geometry with $\theta_{BN}$ = 80 degrees. From the multipoint crossing and shock normal, the velocity of the shock in the spacecraft frame was [-33.5, 23.5, 45.7] ± [2.1, -1.5, -2.9] km/s in GSE, which transforms to [207.5, -1.1, -20.5]$_{GSE}$ km/s in the solar wind rest frame (using the average upstream solar wind velocity of [-241.0, 24.6, 66.2]$_{GSE}$ km/s in the spacecraft frame). From the four-point observations, the shock speed was increasing with an acceleration of ~3 km/s$^2$, which is consistent with the explosive nature of foreshock transients [e.g., Turner et al., 2020]. The propagation speed in the solar wind frame is consistent with this structure being a fast magnetosonic shock, since the estimated Mach numbers for that propagation speed were $M_{Alfvén}$ = 9.9 and $M_{fast}$ = 4.2. Note that MMS was ~5 $R_E$ duskward of the subsolar point of the bow shock at this time, and the nominal orientation of the bow shock surface adjacent to MMS was [0.97, 0.19, 0.13]$_{GSE}$ based on the Fairfield [1971] model. From the bow shock crossings around the time of interest (not shown), MMS's location was in the upstream region of a quasi-parallel oriented bow shock (note, *not* the foreshock transient's shock) and estimated at within 0.5 $R_E$ of the bow shock when the foreshock transient was observed.

Figure 2a shows the relative orientation of the four MMS spacecraft at 04:39:25UT. MMS-2 was located closest to Earth, while MMS-3 was furthest sunward. The four spacecraft were stretched out along the same trajectory with separations ranging from 152 km (MMS-1 to -4) to 723 km (MMS-2 to -3). Those separation scales were comparable to the thermal (and suprathermal) proton gyroradii in the magnetic fields observed around the features of interest: a 2 eV (50 eV) proton with pitch angle of 90-degrees had gyro-radius, $r_{cp}$, of 41, 19, and 10 km (204, 93, and 49 km) in the 5, 11, and 21 nT B-fields around the "foot", "ramp", and "overshoot" features shown around S = 200, 0, and -50 km in Figure 2b, respectively. The corresponding proton gyro-periods were 13, 6, and 3 seconds, respectively. With the spacecraft locations projected onto the shock surface, the maximum separation was 686 km along the shock surface, comparable to the suprathermal $r_{cp}$ in the "foot". Note that foreshock transients, like hot flow anomalies, are on the order of several Earth radii or larger in size [e.g., Turner et al., 2013; Liu et al., 2016], much larger than the MMS separation scales. These are relevant scales to consider for the following analysis and interpretation.

With the shock orientation and speed established, it is possible to convert the time series observed by each MMS spacecraft into a spatial sequence, and considering the geometry of the spacecraft in the system, it is possible to interpret the nature of the observed spatiotemporal structure. Details for the conversion to spatial sequence are included in the appendix. Results of this conversion for |**B**|, density, and current density from MMS are shown in Figure 2, where the distances have been normalized to an origin aligning the features to the initial ramp observed by MMS-3. When distances are not normalized to align the common features, the motion of the trailing edge of the foreshock transient, estimated at ~120 km/s along the shock normal direction (relative to the initial ramp at MMS-3), shifts the features further to the right for each subsequent spacecraft crossing after MMS-3 (see appendix material). Figure 2b shows that each MMS spacecraft observed similar structure during the crossing and highlights the spatiotemporal evolution of the feature at 10 < S < 70 km that rises up and expands to greater S over time (see also Fig. 1h-1k). We refer to that feature at 10 < S < 70 km as the "new shock ramp" structure. With the conversion shown in Figure 2, the original shock ramp was located at S ~ 0 km for all four spacecraft. Key details in Figure 2c-2f include i) large-amplitude B-field waves (note anti-correlation between |**B**| and density) at S < 10 km observed by all four spacecraft; ii) the largely correlated |**B**| and density in the new shock ramp structure observed by all four spacecraft; iii) the ~4x jump in magnitudes of density and |**B**| in the new shock ramp compared to the upstream conditions at S ~ 250 km observed by MMS-1 and -2; iv) oscillations in |**B**| at 30 < S < 160 km observed by MMS-4, -1, and -2; and v) sharp, narrow current density structures concentrated primarily along the sharpest gradients in |**B**| and density and strongest at S = 0 km.

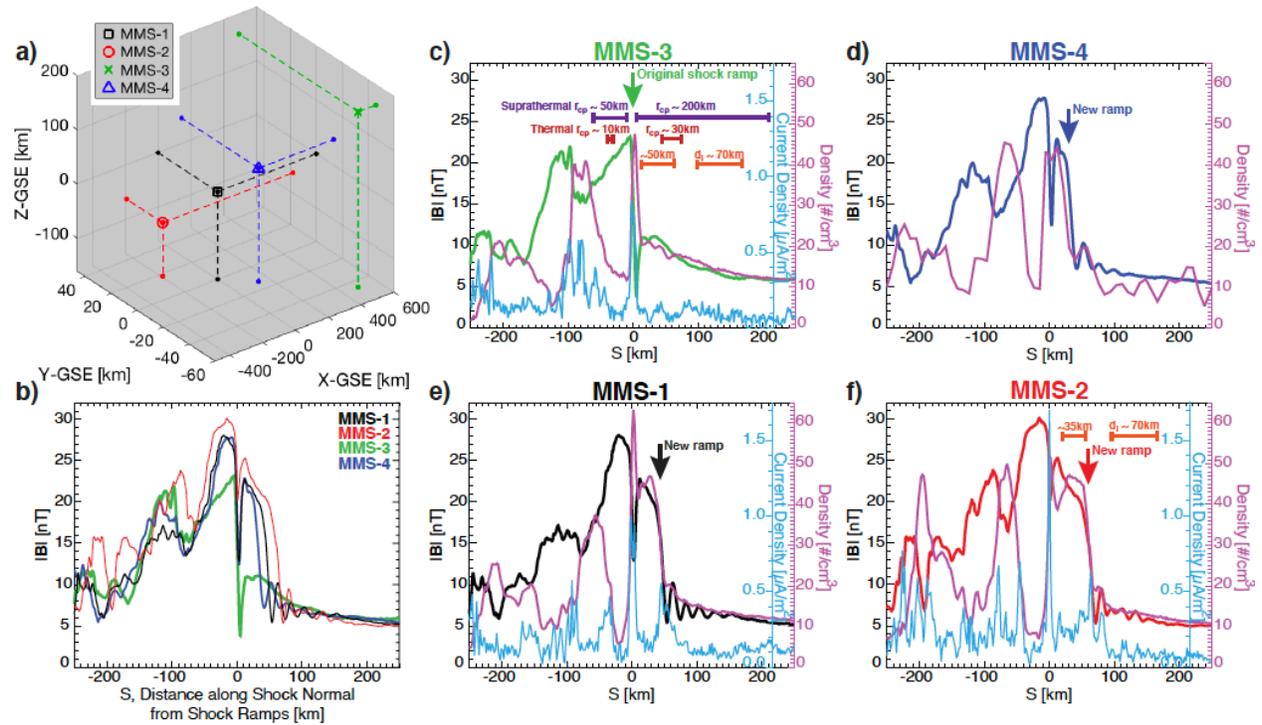

**Figure 2:** a) MMS formation in GSE coordinates centered on MMS-1 location, which was at [14.5, 5.1, 2.5] $R_E$ in GSE at this time. b) Magnetic field magnitudes from all four MMS spacecraft (-1: black, -2: red, -3: green, and -4: blue) plotted along the shock normal direction, S. c) – f) show B-field magnitudes, plasma density, and current density from MMS-3 (c), -4 (d), -1 (e), and -2 (f). B-fields are shown in the respective spacecraft colors, while density and current density are shown in magenta and light blue, respectively. Note that current density is unavailable for MMS-4. The original ramp location is indicated with the green arrow in c), while the new shock ramp locations are indicated with the corresponding colored arrows for MMS-4, -1, and -2 in d), e), and f), respectively. On panel c), examples of thermal (2 eV, dark red) and suprathermal (50 eV, purple) proton gyro-radii are shown on the upstream (S > 0) and downstream (S < 0) regimes, as are examples of the ion inertial length scales (orange) in the upstream regime. Example ion inertial length scales are also shown in the upstream and downstream regimes in f).

Considering the highly correlated nature and time-sequential growth of the feature referred to as the "new shock ramp" observed in sequence by MMS-3, -4, -1, and -2 during their crossings of this shock, it is highly unlikely that the feature was simply the result of random fluctuations along the 3D shock surface. However, we must consider the possibility that it was a coherent structure such as a shock surface ripple [e.g., Lowe and Burgess, 2003; Johlander et al., 2016; Gingell et al., 2017]. Shock ripples reported along Earth's bow shock have wavelengths of ~100 to ~200 km and propagate along the shock surface at speeds of ~65 to ~150 km/s [Johlander et al., 2016; Gingell et al., 2017]. Those wavelengths are comparable to the interspacecraft separation of each adjacent pair of MMS spacecraft in this event. Assuming comparable propagation speeds, a shock surface ripple would pass between MMS-3 and -2 in ~5 to 11 seconds (if propagating perfectly along the inter-spacecraft separation vector) or longer (for different propagation directions). The observed timing between the "new shock ramp" observed at MMS-3 to MMS-2 was ~5 seconds (see Fig. 1), however, Lowe and Burgess [2003] and Johlander et al. [2016] also describe how such surface ripples propagate along the shock surface and parallel to the upstream magnetic field. Considering that propagation angle, it would have taken between ~13 and 30 seconds for a surface ripple to pass from MMS-3 to -2, much longer than the observed ~5-seconds. The ripple explanation becomes even more unlikely when multiple phase fronts propagating along the shock surface are considered. Thus, it is unlikely that the observed feature in question was either random fluctuations or coherent shock surface ripples.

Interpreting the observed "new shock ramp" as resulting from shock reformation, it is possible to use the multipoint MMS observations to calculate the growth rate of the new shock ramp. Using the tangential component of the magnetic field at the overshoots and the shock speed (see also the appendix), we found a shock growth rate of 1.63 nT/s (0.026 nT/km) for the old shock, and 2.55 nT/s (0.041 nT/km) for the reforming shock. A faster growth rate of the new, reforming shock is largely driven by nonlinear steepened waves. These rates have important implications in constraining numerical simulations, which tend to yield unrealistic estimates of reformation rates [e.g., Krasnoselskikh et al., 2013; Scholer, Hinohara, and Matsukiyo, 2003].

The large-amplitude waves observed on the downstream side (S < 0 km) had wavelengths along S comparable to the suprathermal $r_{cp}$ in this frame, and they intensified in amplitude closer to S = 0 km. Around S = 0 ± 10 km, the waves were on electron scales (< 1 ion inertial length, $d_i$) and associated with the intense and thin current layer. Approximately 1 $r_{cp}$ (thermal) upstream of that current layer, around S = 30 km, was where the new shock ramp actually formed. The new shock ramp structure rose up out of the "foot" structure observed by MMS-3 between 10 < S < 200 km, corresponding to within a few thermal $r_{cp}$ upstream of the steepened, electron-scale waves and intense current layer. The new shock ramp itself was observed by MMS-4 first at a scale of ~1 $d_i$ and then growing to ~2 $d_i$ along S by MMS-2. Once the new shock ramp formed, at MMS-1 and -2 in particular, new or intensified electron-scale compressional waves were observed between 0 < S < 30 km, and large-amplitude whistler precursor waves [e.g., Wilson et al., 2012] were observed by MMS-4, -1, and -2 just upstream of the new shock ramp at 30 < S < 160 km. Note those whistler precursors were not observed by MMS-3. The whistler precursor waves were limited to within ~1 $d_i$ upstream of the new shock ramp and exhibited wavelengths ~20 km (i.e., < 1 $d_i$) along S in this frame.

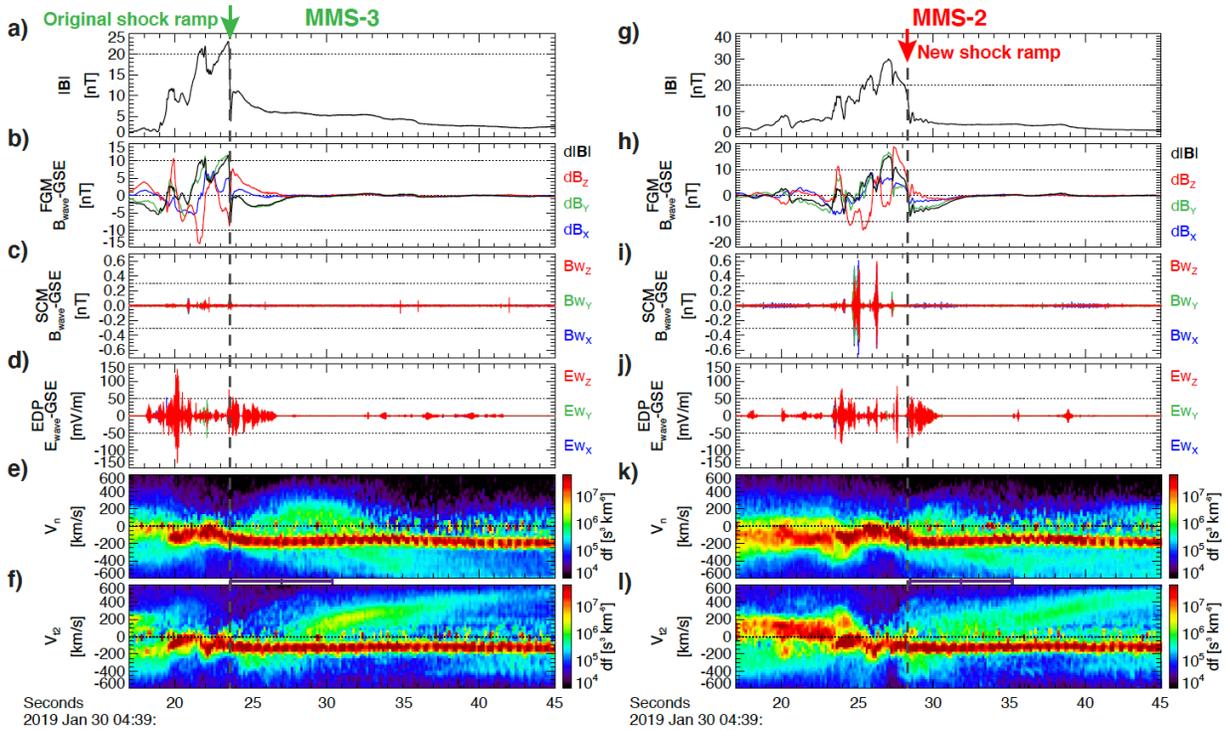

**Figure 3:** Summary of waves and derived data from MMS-3 (a – f) and -2 (g – l). For each spacecraft, the following data are plotted: a) and g) show B-field magnitude (for ease of comparison with other figures); b) and h) show low-frequency $B_{wave}$ ($dB_i = B_i - <B_i>$) from the fluxgate magnetometer data in GSE coordinates (dB-XYZ in blue, green, red, respectively) and d|**B**| in black; c) and i) high-frequency $B_{wave}$ from the search-coil magnetometer data in GSE coordinates; d) and j) high-frequency $E_{wave}$ data from the axial and spin-plane double probe data; e) and k) ion velocity distributions along the shock normal direction in the shock rest frame; f) and l) ion velocity distributions along a vector perpendicular to the shock normal direction in the shock rest frame, highlighting the incident solar wind beam and reflected ion gyration. Note, several of the corresponding plots for MMS-2 and -3 are on different Y-scales, so horizontal dashed lines have been put at the same fixed values on both for ease of comparison.

Figure 3 provides an overview of the electromagnetic and electrostatic waves and reflected ions observed by MMS during this event. Ion acoustic waves were present upstream of the shock observed by all four s/c (-1 and -4 not shown in Fig. 3) after ~04:39:29UT, at which point MMS-3 was too far upstream to determine whether the waves were also present before the new shock ramp formed. Strong broadband electrostatic fluctuations, corresponding to electron-scale nonlinear waves/structures, were observed by all four spacecraft, mostly at gradients in B throughout the downstream regime, particularly near the boundaries at the new shock ramp and edge of the HFA core (~04:29:19UT at MMS-3). The nonlinear waves/structures did not occur simultaneously with the intense, electron-scale current sheets in the downstream regime. The electrostatic nonlinear waves/structures at MMS-3 extended further upstream corresponding with the "foot" structure, whereas for MMS-4, -1, and -2, the fluctuations were limited to approximately the same range in the upstream as the whistler precursors, i.e., within ~1 $d_i$ of the new shock ramp. In the region of the new shock ramp, the amplitude of the electrostatic nonlinear waves/structures was smallest at MMS-3 and largest at MMS-2. The largest amplitude electrostatic waves/structures, >100 mV/m, likely corresponded to very short wavelengths (< 200 m, i.e., less than the tip-to-tip boom length of the spin-plane electric field instruments), which is consistent with observed wavelengths in the shock frame of ~80 – 100 m. Those >100 mV/m waves were only observed in the downstream region, S < 0 km, by MMS-3 and -4, not by -1 and -2. Electromagnetic "lion roars" [e.g., Giagkiozis et al., 2018] were observed in the downstream regime by all four spacecraft, though the amplitude of those whistler mode waves increased significantly after the formation of the new shock ramp; MMS-3 observed lion roars with amplitudes < 100 pT (e.g., around 04:39:20.8UT in Fig. 3c), while MMS-2 observed lion roars at amplitudes > 500 pT (e.g., around 04:39:25.1UT in Fig. 3i). Most interestingly, only at MMS-2 were the lion roars also associated with electrostatic solitary waves (ESWs; examples of which are shown in the supporting material), which is important since such nonlinear wave decay represents a distinctly irreversible energy dissipation process [e.g., Kellogg et al., 2011]. In the shock frame, those ESWs had wavelengths on the order of 100 – 120 m along S, approximately one quarter of the lion roars' wavelengths at ~460 m along S.

Figure 3 also shows ion velocity spectra plotted vs. the shock normal ($V_n$) and tangential ($V_{t2}$) velocity components [e.g., Madanian et al., 2020]. Note that $V_{t2}$ is by definition perpendicular to the shock normal and upstream B-field vectors. The incident solar wind beam is the high-density population at $V_n$ and $V_{t2}$ < 0. The $V_{t2}$ distributions clearly show the energy dispersion effect of ions accelerating and reflecting at the shock ramp: the peak in $V_{t2}$ > 0 ions corresponds to higher energy (larger $V_{t2}$) ions completing a half-gyration (after reflection from the ramp in |**B**|) at increasingly greater distances upstream of the shock. This was true for all four spacecraft (see Fig. 3f and 3l for MMS-3 and -2, respectively), indicating that the shock continues to reflect and accelerate suprathermal ions throughout the reformation process. Note also the differences in $V_n$ from MMS-3 (more intense suprathermal ions at $V_n$ > 0 around 04:39:30UT, corresponding to ~1 suprathermal $r_{cp}$ from the original shock ramp, in Fig. 3f) to MMS-2 (more intense suprathermal ions at $V_n$ < 0 around 04:39:35UT, corresponding to ~1 suprathermal $r_{cp}$ from the new shock ramp, in Fig. 3k), which are possibly cyclical differences coinciding with the different observed phases of the shock reformation cycle. Those distributions include a superposition of ions reflected from the transient structure's shock and the main bow shock plus the incident solar wind, and generation of upstream, ion-scale waves can be associated with any of these populations plus interactions between them.

## 4. Summary and Conclusion

At 04:39 UT on 30 Jan 2019, MMS was fortuitously positioned to capture what was likely at least half of the reformation cycle of a fast magnetosonic shock on the upstream edge of a transient structure in the quasi-parallel foreshock upstream of Earth's bow shock. Evidence was provided supporting that it was unlikely that the observed features resulted from either random fluctuations or shock surface ripples when the spacecraft separation tangential to the shock normal was also accounted for. This unique case study offered an opportunity to study the spatiotemporal nature of early shock development in microscopic detail. Calculated shock growth rates indicated that the new shock ramp grew faster (2.55 nT/s) than the old shock ramp (1.63 nT/s). As the new shock ramp formed from the "foot" of the pre-existing shock, several additional distinct differences were observed down to electron kinetic scales, including intensification of electron-scale waves, nonlinear waves/structures, and intense current sheets. It was at those electron kinetic scales ($< \sim 1\ d_i$) that the new shock ramp first formed before expanding back up into the ion scales ($> \sim 1\ d_i$). Prior to the shock ramp reforming, the steepened, large-amplitude ion-scale wavefronts were also affecting electrons, resulting in the growth of electrostatic and electromagnetic wave modes and thin, intense current layers. However, once the new shock ramp was properly established, as exemplified by MMS-2, both the electrostatic and electromagnetic waves amplified significantly at the new shock ramp and in the downstream region. The most intense current layer was observed along the original shock ramp (around S = 0 km), and the new shock ramp and an overshoot formed immediately upstream and downstream of that intense, electron-scale current layer, respectively. Note that the overshoot on the downstream side of the current layer was likely that of the original shock ramp, and from the available snapshots of the new ramp, it is difficult to identify where any new overshoot was formed. Only after the new shock ramp formed were whistler precursors in the upstream region and potentially dissipative wave-wave interactions in the downstream region observed. All combined, the results indicate that a shock's energy conversion and dissipation processes may also undergo the same cyclical periodicity as reformation of the shock front.

This special case exemplifies the genuine cross-scale coupling that occurs between the ion- and electron-kinetic physics at collisionless, fast magnetosonic shocks. The ions, with their large gyro-radii, enable information transfer "very far" (with respect to electron scales) into both the upstream and downstream regimes, but the key physics for energy dissipation and heating occur at least in some relevant part at electron scales via thin, intense, electron-scale current sheets and large-amplitude, nonlinear electrostatic fluctuations and electromagnetic (e.g., whistler precursors just upstream and lion roars throughout the downstream) waves. Throughout the reformation cycle, the enhanced |**B**| at the ramp, overshoot, and downstream reflects a significant fraction of incident solar wind ions back into the upstream regime, resulting in the development of the diamagnetic "foot"-like structure, out of which the new shock ramp formed. During the reformation process before the new ramp forms, ion-scale waves steepen and compress in what will ultimately become the new downstream regime. Critically, the compression of the waves reaches electron-kinetic scales, where strong energy transfer then begins along thin, intense current sheets and in the large-amplitude, electron-kinetic-scale waves. The compressed waves and current-sheet energy transfer at electron-scales culminate in the formation of a new shock ramp, with correlated |**B**| and density, out of the pre-existing "foot"-like structure upstream of the most-intense, thin current layer. Once formed, the new shock ramp and "foot" region continue converting energy of the incident ion and electron populations via whistler-mode precursor and electrostatic fluctuations within a few $d_i$ upstream of the shock ramp, dissipative wave-mode-coupling downstream of the ramp, and along thin current layers that may also be reconnecting [e.g., Gingell et al., 2019; Liu et al., 2020]. As we know from many observations of foreshock transient shocks, the extent of the shocked plasma then must expand rapidly back up to ion-kinetic and ultimately MHD scales.

## Acknowledgments
The authors are thankful to the MMS team for making their data available to the public. We thank the ACE, Wind, and OMNI teams and data providers for solar wind data. We are also thankful to the anonymous reviewer for constructive suggestions and comments. Funding support for several authors was via the MMS mission, under NASA contract NNG04EB99C, and research supported by the International Space Science Institute's International Teams program. DLT is also thankful for funding from NASA grants (NNX16AQ50G and 80NSSC19K1125). SJS was supported in part by a subcontract from Univ. of New Hampshire on NASA award 80NSSC19K0849. HM was supported in part by NASA grant 80NSSC18K1366. HH was supported by the Royal Society University Research Fellowship URF\R1\180671. The French LPP involvement for the SCM instrument was supported by CNES and CNRS. SPEDAS software [Angelopoulos et al., 2019] was used to access, process, and analyze the MMS data. MMS data used in this study are available at https://lasp.colorado.edu/mms/sdc/public/.

## Figure Captions

**Figure 1:** Overview of the event observed by MMS. a) – g) show data from the foreshock transient observed by MMS-1 on 30 Jan 2019, including: a) magnetic field vector in GSE coordinates (XYZ in blue, green, and red, respectively) and magnitude (black); b) ion omnidirectional energy-flux (color, units eV/cm$^2$-s-sr-eV); c) electron density; d) ion velocity vector in GSE coordinates (XYZ in blue, green, and red, respectively) and magnitude (black); e) proton gyro-radius (color) as a function of energy and time; f) proton gyro-frequency; g) ion inertial length. h) – k) show magnetic field vectors and magnitudes from all four MMS spacecraft zoomed in on the feature of interest in this study.

**Figure 2:** a) MMS formation in GSE coordinates centered on MMS-1 location, which was at [14.5, 5.1, 2.5] $R_E$ in GSE at this time. b) Magnetic field magnitudes from all four MMS spacecraft (-1: black, -2: red, -3: green, and -4: blue) plotted along the shock normal direction, S. c) – f) show B-field magnitudes, plasma density, and current density from MMS-3 (c), -4 (d), -1 (e), and -2 (f). B-fields are shown in the respective spacecraft colors, while density and current density are shown in magenta and light blue, respectively. Note that current density is unavailable for MMS-4. The original ramp location is indicated with the green arrow in c), while the new shock ramp locations are indicated with the corresponding colored arrows for MMS-4, -1, and -2 in d), e), and f), respectively. On panel c), examples of thermal (2 eV, dark red) and suprathermal (50 eV, purple) proton gyro-radii are shown on the upstream (S > 0) and downstream (S < 0) regimes, as are examples of the ion inertial length scales (orange) in the upstream regime. Example ion inertial length scales are also shown in the upstream and downstream regimes in f).

**Figure 3:** Summary of waves and derived data from MMS-3 (a – f) and -2 (g – l). For each spacecraft, the following data are plotted: a) and g) show B-field magnitude (for ease of comparison with other figures); b) and h) show low-frequency $B_{wave}$ ($dB_i = B_i - <B_i>$) from the fluxgate magnetometer data in GSE coordinates (dB-XYZ in blue, green, red, respectively) and d|**B**| in black; c) and i) high-frequency $B_{wave}$ from the search-coil magnetometer data in GSE coordinates; d) and j) high-frequency $E_{wave}$ data from the axial and spin-plane double probe data; e) and k) ion velocity distributions along the shock normal direction in the shock rest frame; f) and l) ion velocity distributions along a vector perpendicular to the shock normal direction in the shock rest frame, highlighting the incident solar wind beam and reflected ion gyration. Note, several of the corresponding plots for MMS-2 and -3 are on different Y-scales, so horizontal dashed lines have been put at the same fixed values on both for ease of comparison.

## Appendix: Supporting Material

**Calculating the Local Bow Shock Orientation:**
Local bow shock normal direction from the Fairfield [1971] model:
$$\mathbf{n_{bs}} = [0.974, 0.190, 0.127] \text{ in GSE}$$
Upstream magnetic field (average from MMS-1):
$$\mathbf{B} = [2.00, 2.37, 0.35] \text{ nT in GSE}$$
Angle between bow shock normal and upstream B-field:
$$\theta_{Bn} = 38.5 \text{ degrees}$$
So, MMS were in the quasi-parallel foreshock, consistent with plasma and field observations and the presence of the foreshock transient structure.

**Calculating the Foreshock Transient's Shock Normal Direction and Orientation:**
Using coplanarity with B and V from Schwartz [1998]:
$$\mathbf{n} = (\Delta\mathbf{B} \times \Delta\mathbf{V} \times \Delta\mathbf{B})/|(\Delta\mathbf{B} \times \Delta\mathbf{V} \times \Delta\mathbf{B})|$$
where $\Delta\mathbf{X} = \mathbf{X}_{downstream} - \mathbf{X}_{upstream}$, $\mathbf{X} = \mathbf{B}$ or $\mathbf{V}$

| Coplanarity, B and V: | Downstream Times: 04:UT+ | Upstream Times: 04:UT+ | n-GSE |
|---|---|---|---|
| MMS-1 | 39:24.0 - 39:26.0 | 39:30.0 - 39:32.0 | [0.389, -0.315, -0.866] |
| MMS-2 | 39:25.0 - 39:27.0 | 39:31.0 - 39:33.0 | [0.590, -0.551, -0.620] |
| MMS-3 | 39:20.5 - 39:23.0 | 39:27.0 - 39:29.0 | [0.670, -0.338, -0.661] |
| MMS-4 | 39:22.5 - 39:25.3 | 39:28.0 - 39:30.0 | [0.512, -0.313, -0.800] |

Average **n** from all four s/c ± 1 standard deviation on the mean:
$$\mathbf{n_{sh}} = [0.540, -0.379, -0.737] \pm [0.104, 0.010, 0.100] \text{ in GSE}$$

Magnetic field upstream of the transient's shock:
$$\mathbf{B} = [1.94, 1.16, 0.30] \text{ nT in GSE}$$
Angle between transient shock normal and upstream B-field:
$$\theta_{Bn} = 80.3 \text{ degrees}$$
So, the transient shock was in a quasi-perpendicular orientation

**Calculating the Foreshock Transient's Shock Speed:**

| Spacecraft: | Shock Ramp t | Location in GSE [km] | Velocity [km/s] |
|---|---|---|---|
| MMS-1 | 04:39:26.320 UT | [92772.145, 32264.144, 16086.952] | [-1.715, 0.155, -0.562] |
| MMS-2 | 04:39:27.310 UT | [92579.885, 32282.023, 16023.796] | [-1.719, 0.154, -0.563] |
| MMS-3 | 04:39:23.600 UT | [93264.432, 32218.352, 16248.785] | [-1.704, 0.159, -0.560] |
| MMS-4 | 04:39:25.550 UT | [92915.644, 32250.873, 16134.104] | [-1.712, 0.156, -0.561] |

Shock speed: $V_{sh} * \Delta t = \Delta \mathbf{X} \cdot \mathbf{n}_{sh}$

| Spacecraft Pairs: | $|\Delta X|$ [km] | $\Delta t$ [s] | Shock Speed [km/s] |
|---|---|---|---|
| MMS-3 to -4 | 368.6 | 1.94 | -59.9 |
| MMS-3 to -1 | 520.2 | 2.72 | -60.5 |
| MMS-3 to -2 | 723.4 | 3.70 | -61.7 |
| MMS-4 to -1 | 151.6 | 0.78 | -61.7 |
| MMS-4 to -2 | 354.8 | 1.79 | -63.6 |
| MMS-1 to -2 | 203.2 | 0.99 | -65.1 |

Average shock speed along shock normal ± 1 standard deviation:
$$V_{sh} = -62.1 \pm 1.9 \text{ km/s}$$
Note, the shock is apparently accelerating along the shock normal direction at an average rate of:
$$a = 2.93 \text{ km/s}^2$$

Shock velocity in spacecraft frame (GSE):
$$\mathbf{V}_{sh}|sc = [-33.5, 23.5, 45.7] \pm [2.1, -1.5, -2.9] \text{ km/s}$$

Shock velocity in solar wind frame (GSE):
$$\mathbf{V}_{sh}|sw = [207.5, -1.1, -20.5] \pm [2.1, -1.5, -2.9] \text{ km/s}$$

Mach Numbers in background solar wind:
$$M_{Alfvén} = 9.9, M_{fast} = 4.2$$

**Converting MMS Time Series to Distance Along Shock Normal Vector:**

$\mathbf{V}_{MMS}|sh = \mathbf{V}_{MMS} - \mathbf{V}_{sh}$ : MMS velocity in shock reference frame
$v_{MMS}|sh = \mathbf{V}_{MMS}|sh \cdot \mathbf{n}_{sh}$ : MMS apparent speed along shock normal
$t_3(0)$ : time at which MMS-3 observed the original shock ramp (see "Shock Ramp t" in table above)
$\Delta t_i = t_i(t) - t_i(0), i = \{1, 2, 3, 4\}$
$\Delta S_i = v_{MMS}|sh * \Delta t_i$

For each spacecraft, $i = \{1, 2, 3, 4\}$, $\Delta S_i$ is then the distance along the shock normal direction from the original shock ramp location; thus, $S = 0$ is where each MMS spacecraft first observed the original shock ramp.

Note that for the results shown in Figure 2 of the paper, S were calculated using $t_i(0)$ for each spacecraft. See Figure SM1 below for example of the spatial series plotted vs. S where all four spacecraft are referenced to the location (S=0) of the original shock ramp when/where it was first observed by MMS-3 at $t_3(0)$. That conversion showcases the expansion speed of the foreshock transient but does not align common features between all four spacecraft.

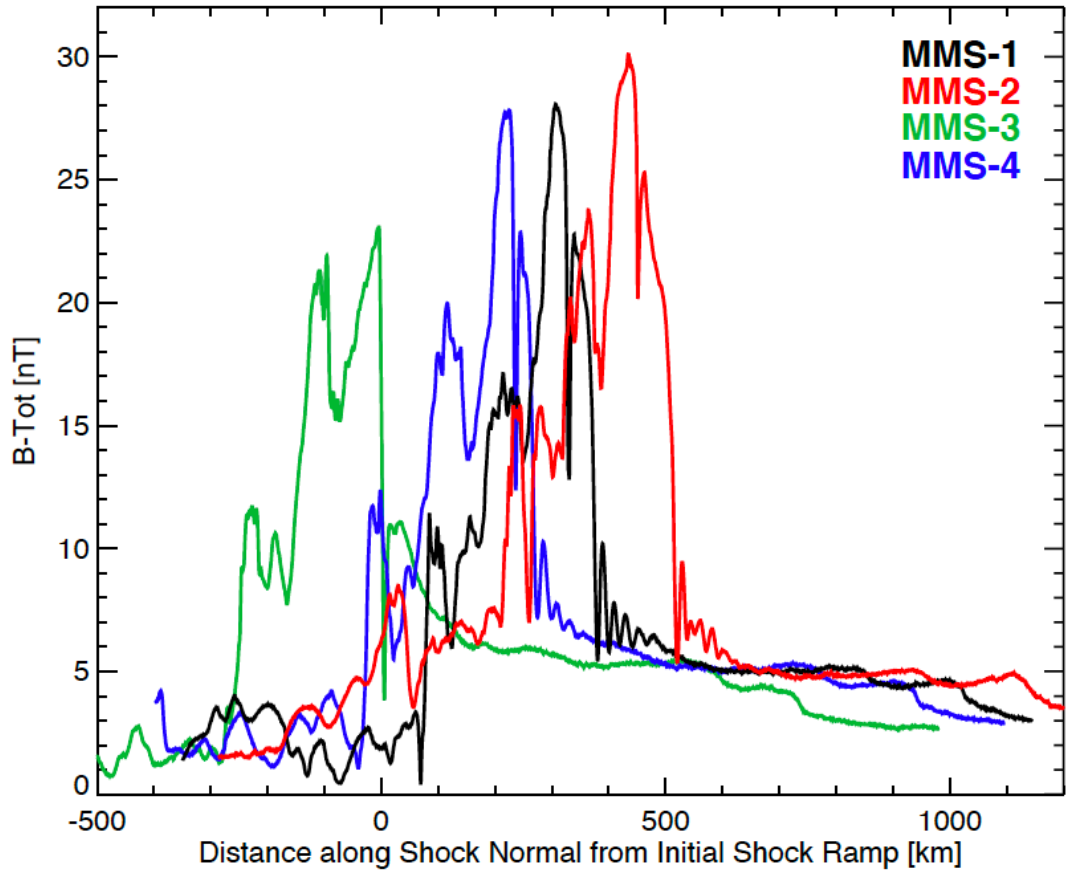

**Figure SM1:** |B| time series from each MMS spacecraft converted to distance along shock normal direction using $t_3(0)$ for all four spacecraft, i.e., instead of $t_i(0)$. This conversion captures the expansion of the foreshock transient but does not align common features along this version of S.

**Foreshock Transient Shock Orientation Sketches:**

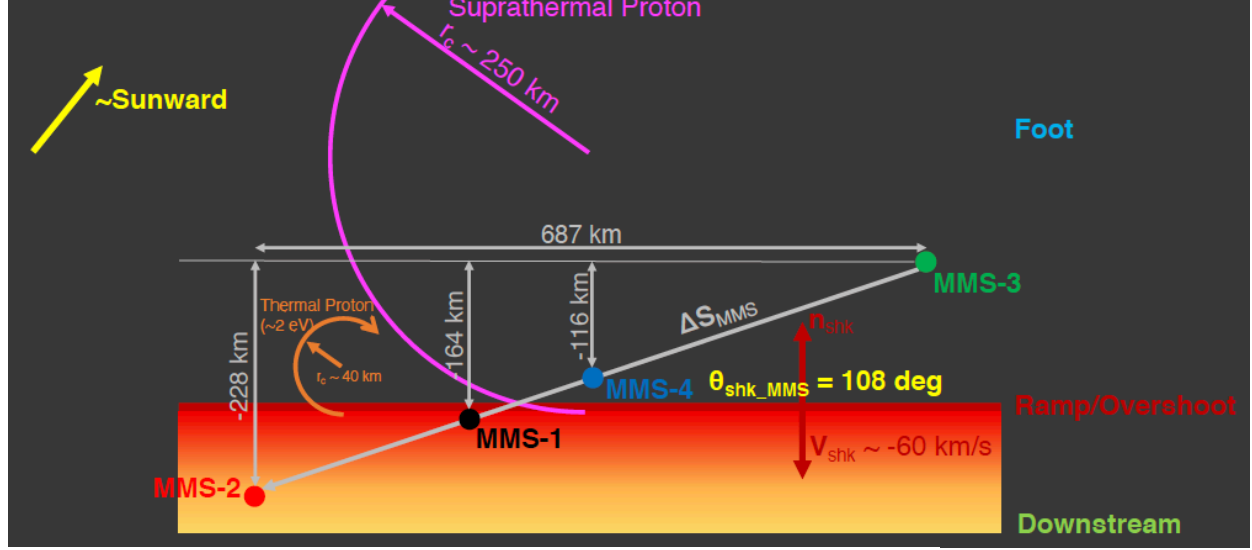

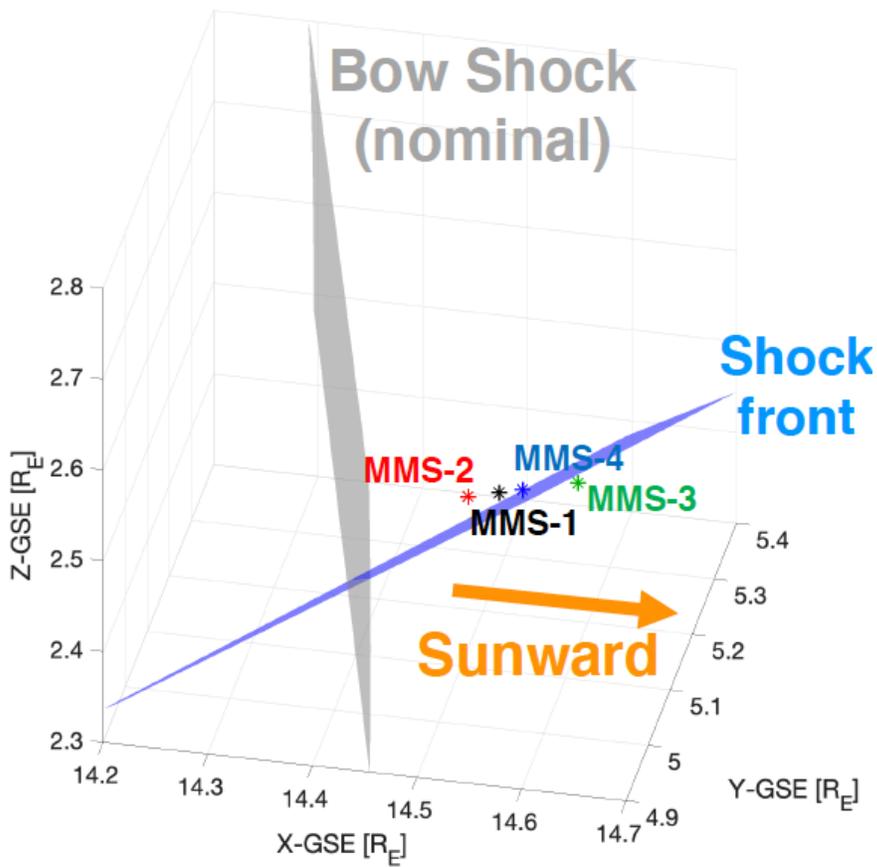

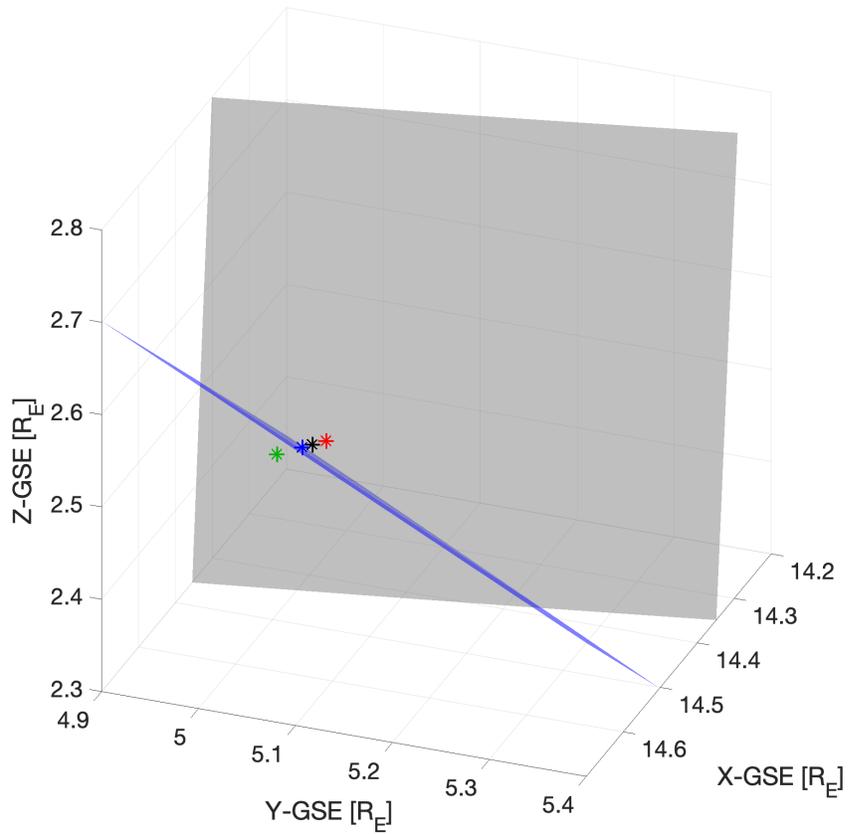

**Figure SM2:** Sketches of the orientation and relative size scales of the foreshock transient shock and MMS spacecraft and the bow shock.

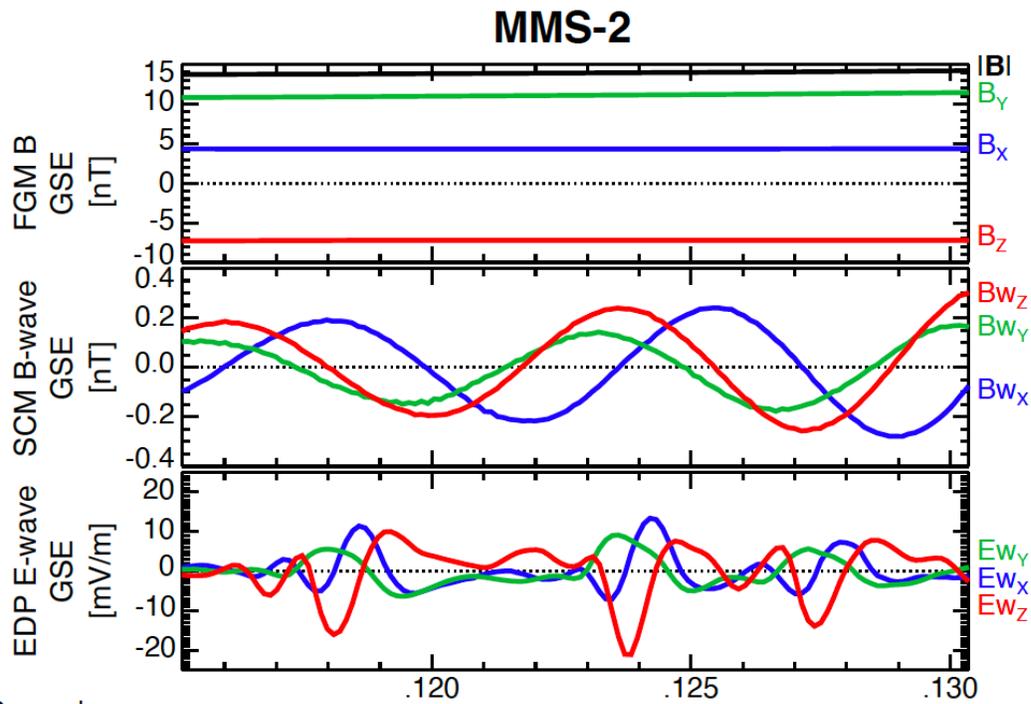

**Figure SM3:** Example of possible nonlinear wave decay on electron-kinetic-scales. From top to bottom, the three panels show top: magnetic field vector in GSE (XYZ in blue, green, and red) and magnitude (black); mid: $B_{wave}$ in GSE coordinates from the search-coil magnetometer; bot: $E_{wave}$ in GSE coordinates from the electric field double probes. The middle panel shows approximately two wavelengths from an electromagnetic whistler-mode "lion roar" observed by MMS-2 in the downstream plasma regime, while the bottom panel shows three, large-amplitude electrostatic solitary waves.